\documentclass{PoS}

\title{Using Feynman's Tree Theorem to Evaluate Loop Integrals Numerically}

\ShortTitle{Using Feynman's Tree Theorem to Evaluate Loop Integrals Numerically}

\author{Wolfgang Kilian\\
        University of Siegen, Fachbereich Physik\\
        E-mail: \email{kilian@physik.uni-siegen.de}}
        
\author{\speaker{Tobias Kleinschmidt}
         \thanks{Supported by the UK Science and Technology Facilities Council (STFC), PP/E 0073717/1.}\\
        University of Durham, Institute for Particle Physics Phenomenology\\
        E-mail: \email{tobias.kleinschmidt@durham.ac.uk}}

\abstract{We report on a new method for the numerical evaluation of loop
  integrals, based on the Feynman Tree Theorem.  The
  loop integrals are replaced by phase-space integration over
  fictitious extra on-shell particles.  This integration can be
  performed alongside with the Monte-Carlo integration of ordinary
  phase space,  directly leading to NLO event generation.
  }

\FullConference{RADCOR 2009 - 9th International Symposium on Radiative Corrections (Applications of Quantum Field Theory to Phenomenology) \\
                 October 25-30 2009\\
                 Ascona, Switzerland}

\newcommand{\feynarts}{\texttt{FeynArts}}
\newcommand{\formcalc}{\texttt{FormCalc}}
\newcommand{\vamp}{\texttt{VAMP}}
\newcommand{\GeV}{{\ensuremath\rm GeV}}

\begin{document}

\section{Introduction}
A crucial problem of event generation at next-to-leading order (NLO) in
perturbation theory originates from the loop
contributions to multi-parton matrix elements. The number of
individual Feynman graphs rises dramatically with the number of
external legs, and tensor reduction methods increase the number of
terms even more.  This poor scaling behavior, and the associated
instabilities due to large numerical cancellations in matrix elements,
make it worthwhile to investigate different and alternative approaches
to the problem of automatic NLO computation and simulation.

We propose a new method for the evaluation of loop integrals, which allows for
direct numerical computation without a decomposition into a set of basic integrals
and corresponding large coefficients.\footnote{
This is in contrast to methods used for recent computations of $2\rightarrow 4$
processes at the LHC~\cite{Automated:NLO}. These either use Passarino-Veltman
style reductions or a unitarity based approach to decompose amplitudes
into a set of basic integrals.} These proceedings summarize the work of a
recent publication~\cite{Kilian:2009wy}. Here, the matrix element is re-expressed using an
improved version of the Feynman Tree Theorem
(FTT)~\cite{FTT}. The loop integrals are transformed into
phase space integrals and are evaluated numerically along with the
phase space integral over the external partons of the process under
consideration.  Thus, there is only one step of numerical integration
involved in the computation of any particular integrated cross section
or distribution.  We therefore can make use of powerful existing
technologies for numerical phase-space integration by tree-level event
generators to evaluate NLO processes.

\section{Feynman Tree Theorem}\label{s_ftt}

The integrand $I(k)$ of a loop integral over momentum $k$ can be written as a product of
Feynman propagators $F$  times a regular function $N(k)$ in the numerator.
Suppressing possible indices in the numerator, we have
\begin{equation}\label{integrand}
I(k)=N(k)\prod_{i}F(k+p_i,m_i).
\end{equation} 
The $p_i$ are linear combinations of external momenta,  $m_i$
the masses of the physical particle the propagator corresponds to.
We define
\begin{equation}\label{fprop}
F_i\equiv F(k+p_i,m_i)
=\frac{i}{(k+p_i)^2-m_i^2+i\epsilon},
\end{equation}
using 't Hooft-Feynman gauge throughout this discussion.

The Feynman Tree Theorem~\cite{FTT} states, that the loop integral over
(\ref{integrand}) can be replaced by an integral over
a sum of terms, where in each term one or more propagators $F_i$ are
replaced by $\delta$-functions,
\begin{equation}
\int \! dk \cdot I(k)=\!\! \int \! dk \cdot N(k)\left[\sum \Delta^l F \cdots -\sum \Delta^l \Delta^l F \cdots 
+\ldots -(-1)^{n} \sum \Delta^l\cdots\Delta^l\right].\label{ftt_original}
\end{equation}
with $dk\equiv d^4k\cdot (2\pi)^{-4}$. The $\delta$-functions 
$\Delta_i^l=\frac{2\pi}{2E_i}\delta(k^0-(-p^0_i+E_i))$ cancel the $k^0$-integration 
and cut the loop by setting the momentum of the  original propagator $F_i$ 
on its mass-shell. The numerator structure of this propagator factorizes into a
product of wave functions of the corresponding particle. The original loop integral
can therefore be written as a sum of tree graphs, with a phase space
integration replacing the loop integral.

Derivations of the FTT in its form (\ref{ftt_original}) can also be found 
in~\cite{Brandhuber:2005kd} and~\cite{Catani:2008xa}. In the latter work,
Catani \textit{et  al.} exploited the relation between loop integrals and 
phase space integrals to express loop amplitudes as sum of tree amplitudes
arising solely from single cuts.  Contributions from multiple cuts, which are
present in (\ref{ftt_original}), are compensated by a non-trivial $i\epsilon$-prescription
in the propagators.

To make use of the FTT in a numerical evaluation of loop integrals
involving only real numbers, at some point we have to set the
$i\epsilon$ terms in the denominators to zero. Using the identity
\begin{equation}\label{principal_value}
\frac{1}{x-a\pm i\epsilon}=\mathcal{P}\frac{1}{x-a}\mp i\pi\delta(x-a),
\end{equation}
\noindent where $\mathcal{P}$ is Cauchy's Principal Value, we can
rewrite (\ref{fprop}) after a partial fraction
decomposition:
\begin{equation}\label{ftop}
F_i = P_i+\frac{1}{2}\Delta^l_i+\frac{1}{2}\Delta^u_i ,\label{P}
\end{equation}
\noindent where we defined $\Delta^u_i=\frac{2\pi}{2E_i}\delta(k^0-(-p^0_i-E_i))$, as the delta function that
sets the associated four-momentum on its mass-shell with negative zero component. $P_i$ stands for the
propagator with no $i\epsilon$-prescription in the denominator.

Replacing the propagators $F_i$ in (\ref{ftt_original}) by (\ref{ftop}), we obtain a version
of the FTT, which is better suited for a direct numerical integration:
\begin{eqnarray}
\int\! dk \cdot I(k)\!\!\!&=\!\!\!&\int dk\cdot N(k)\big[
\Delta^l_1 P_2\cdots P_n + P_1\Delta^l_2 P_3\cdots P_n + \ldots + 
P_1\cdots P_{n-1}\Delta^l_n\big]\nonumber\\
&&\vspace{1mm}\nonumber\\
&&+\int\!dk\cdot N(k)\!\!\!\!\!\sum\limits_{\mbox{\tiny $\begin{array}[t]{c}perm.\\U+L\ge 2\end{array}$}}\!\!\!\!\!\! \frac{1}{2^{L+U}}\left(1-(-1)^L\right)\,\,\Delta^{l^L}\Delta^{u^U} P^P.\label{ftt}
\end{eqnarray}
The sum runs over all possible permutations, where the functions
($\Delta^l$,$\Delta^u$,$P$) appear ($L$,$U\!$,$P$) times.

We explicitly wrote out the terms containing one $\Delta^l$ function
in the first line of (\ref{ftt}). After $k^0$-integration, all of these terms can be
interpreted as tree graphs with one additional incoming and outgoing
particle and an additional phase space integral over this particles
momentum. This integral can now be pulled out of the individual graphs and put in front of the
amplitude, alongside the phase space integrals over the external
particles which are present in a cross-section calculation.  Since the
extra integration is also of the form of a phase-space integral,
techniques developed for the integration, in particular multi-channel
sampling, can immediately be adopted.  Furthermore, this integration
can be performed \emph{simultaneously} with the external phase space
integration, which is in contrast to common analytical methods,
where for each individual configuration of external momenta the
analytical result of the loop integration has to be numerically
evaluated. 

If the momenta of two or more propagators go on-shell
simultaneously in the momentum integration, terms with more than one
$\Delta$-function in (\ref{ftt}) are non-vanishing. These give rise to further
contribution to the final result, but can be evaluated rather easily, since after the 
$k^0$-integration, the remaining $\delta$-functions lower the dimension 
of integration to two or less.

\section{UV and IR Subtraction}

As long as we are dealing with massive theories, a convenient set of
renormalization conditions is given by the on-shell renormalization scheme.
This requires the pole of the real part of propagators to coincide with the
corresponding particle mass, with residue~$1$. Vertex functions are set
equal to the tree level vertex functions for on-shell external legs.

We make use of a variation of the BPHZ procedure, which results
in loop graphs acting as counterterms. These  can be evaluated under
the same phase space integral over the loop momentum. UV divergences 
are then cancelled locally in three-momentum space.

Consider a 1PI one-loop graph $\Gamma^n(p_1,\dots,p_n)$ with
superficial degree of divergence $\omega(\Gamma)$. We define the T
operator as a Taylor expansion around on-shell momenta $\bar{p}_i$,
with $\bar{p}_i^2=m_i^2$:
\begin{eqnarray}\label{t_op}
T\circ \Gamma^n(p_1,\dots ,p_n) &=&  \Gamma^n(\bar{p}_1,\dots ,\bar{p}_n)+
\sum\limits_{i}^{n-1}(p_i-\bar{p}_i)^\mu \left. \frac{\partial \Gamma^n}{\partial p_i^\mu}\right|_{p_1=\bar{p}_1,\dots ,p_n=\bar{p}_n}+\nonumber\\&&
\dots +\\ &&
\frac{1}{d!}\sum\limits_{i_1,\dots ,i_d}^{n-1} (p_{i_1}-\bar{p}_{i_1})^{\mu_1} \dots (p_{i_d}-\bar{p}_{i_d})^{\mu_d}\left.\frac{\partial^d \Gamma^n}{\partial p_{i_1}^{\mu_1}\dots\partial p_{i_d}^{\mu_d}}\right|_{p_1=\bar{p}_1,\dots ,p_n=\bar{p}_n}\nonumber,
\end{eqnarray}
\noindent up to $d=\omega(\Gamma)$. With this T operator, the renormalized 1PI n-point functions
\begin{equation}
\hat{\Gamma}^n(p_1,\dots ,p_n) =  \Gamma^n(p_1,\dots ,p_n) - T \circ \Gamma^n(p_1,\dots ,p_n)
\end{equation}
fulfill the renormalization conditions of the on-shell scheme.

The expressions resulting from (\ref{t_op}) can also be interpreted as
loop graphs and are easily derived from the original Feynman graph.
Applying again the FTT on the subtraction graphs, the resulting phase
space integral over all tree graphs is UV finite.

A closer inspection of the tree graphs of the FTT reveals, that the IR divergent
parts solely arise from terms with a cut massless propagator. These can be
immediately related to the corresponding IR divergent real emission graph by
crossing the incoming piece of the cut propagator.

Although in the limit of zero momentum-flow through this particle, the two
contributions cancel each other, a suitable projection of the real emission part
onto the virtual part is needed to cancel the IR peak locally in three-momentum space.
For soft divergences, a simple approximation is obtained by setting the two contributions
equal for momenta lower than a given soft energy $E_s$.

In the light of collinear divergences, a more sophisticated ansatz is to construct dipole
subtractions, which cancel IR divergences locally also on the virtual side. The sum of the
resulting analytic integrals for both parts of the dipoles would then be finite. This is left for
future works.

\section{Threshold Singularities}

When the momentum integration is performed in the first line of
(\ref{ftt}), the integrand might get peaks in parts of the phase space
where momenta of un-cut propagators are on-shell. The occurrence
of such peaks, although analytically integrable, leads to problems
in the numerical evaluation of the integrand. In a series of papers for a
direct numerical integration of massless amplitudes,
Soper \textit{et al.}~\cite{Soper:Nagy} proposed a
contour deformation of the integration into the complex plane to avoid
the singularities. To allow for a direct implementation in an ordinary
Monte Carlo generator, we add subtraction terms with zero
real value but with the same peak structure to the integrand. These
will cancel the peaks and allow for a better convergence in the numerical
evaluation.

In the sum of the tree graphs, peaks remain if
$p^2_{ji}\equiv(p_i-p_j)^2>(m_i+m_j)^2$. The momentum constellation is such,
that the two real particles could be produced at the same time. In the rest frame of $p^2_{ji}$,
the peak of the threshold singularity is spherical with radius ${\bf k}_s$ and a simple pole
\begin{equation}
I({\bf k})\propto ({\bf k}-{\bf k}_s)^{-1}, \hspace{2cm}{\bf k}_s=\lambda^{\frac{1}{2}}({p^0_{ji}}^2,m_i^2,m_j^2)\cdot (2|p^0_{ji}|)^{-1},
\end{equation}
$\lambda$ being the K\"all\'en function. We can now evaluate the residue of the integrand and divide it by the singular
structure. Adding the resulting function to the integrand symmetrically around the spherical peak,
the peak will be cancelled and the integrated contribution of the fix function vanishes. More precisely, 
after a Lorentz transformation into the rest frame of $p^2_{ji}$, the function we use reads
\begin{equation}\label{dff}
\mbox{F}(\mathbf{k'})\!\!\equiv\!\!\frac{\mathbf{k}_s R(\Lambda^{-1}k'_s-p_i)}{4p_{ji}^0}\!\left(\frac{1}{\mathbf{k'}-\mathbf{k}_s}-2\frac{\mathbf{k'}-\mathbf{k}_s}{c^2}+
\frac{(\mathbf{k'}-\mathbf{k}_s)^3}{c^4}\right)\!\!
\theta(\mathbf{k'}-(\mathbf{k}_s-c))\theta((\mathbf{k}_s+c)-\mathbf{k'}).
\end{equation}
This has to be transformed back to the integration frame and added to the integrand.
Here, we also added a linear and cubic term in $(\mathbf{k'}-\mathbf{k}_s)$,
to make the joined integrand continuous and differentiable at the artificial borders
introduced by the theta functions.

It is possible, that for higher n-point functions there are overlapping threshold singularities.
In this case, we can add further fix functions similar to (\ref{dff}) to cancel the peak structure.
Since in this case, some of the added functions do not vanish after integration and
give a small contribution to the final result, the width parameter $c$ in (\ref{dff}) should be
taken considerably small. This leads to a trade-off between accuracy and efficiency
in the final integration.

\section{Results}
\begin{figure}[t]
\begin{center}
\begin{picture}(150,48)
\put(0,0)
{\begin{minipage}[b]{7.5cm}
\includegraphics[width=7.5cm]{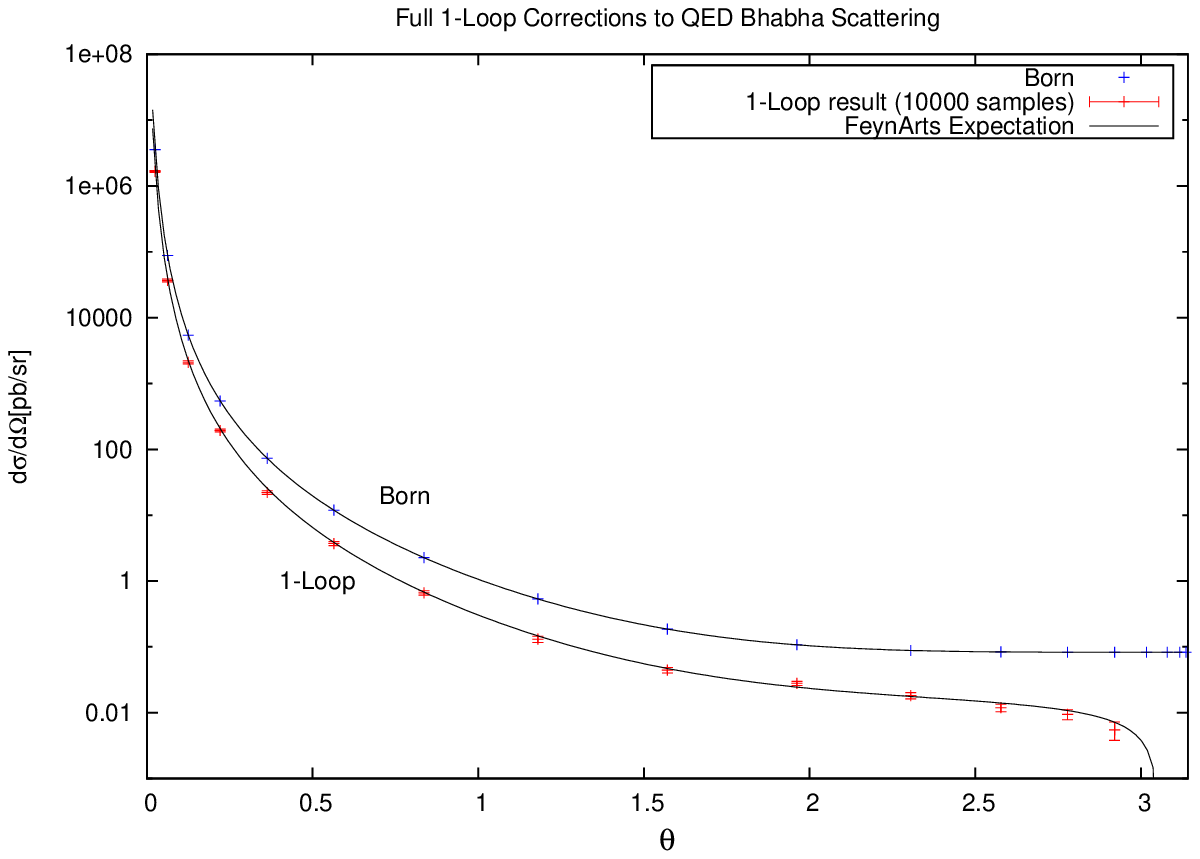}
\end{minipage}}
\put(75,0)
{\begin{minipage}[b]{7.5cm}
\includegraphics[width=7.5cm]{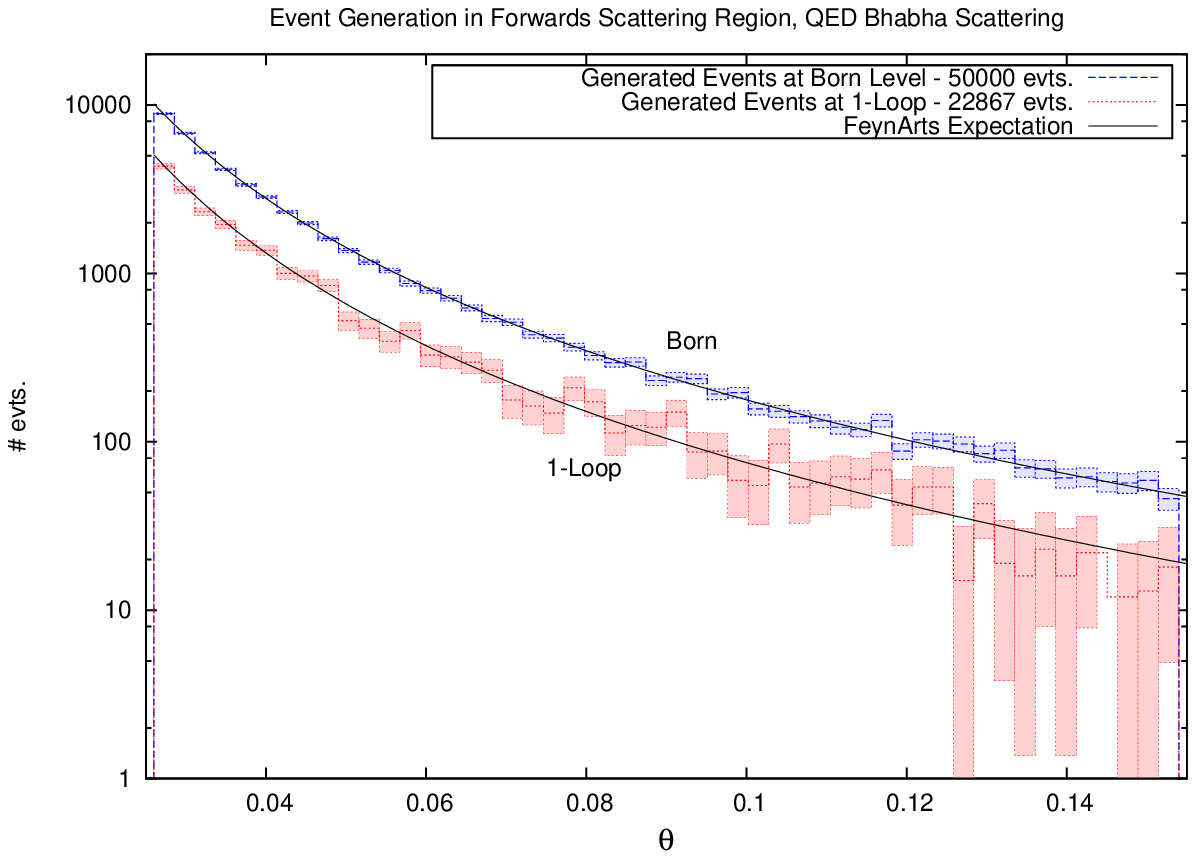}
\end{minipage}}
\put(20,40){$\begin{array}{l} {\scriptscriptstyle \sqrt{s}=500\mbox{{\tiny \GeV}} } \\
	{\scriptscriptstyle \Delta E_s=5\mbox{{\tiny \GeV}}} \end{array}$}
\put(88,13){$\begin{array}{l} {\scriptscriptstyle \sqrt{s}=500\mbox{{\tiny \GeV}} } \\
	{\scriptscriptstyle \Delta E_s=5\mbox{{\tiny \GeV}}} \end{array}$}
\end{picture}
\vspace{-3mm}
\caption{Differential cross sections for Monte Carlo integration and event generation of Bhabha scattering.} \label{f_results}
\end{center}
\vspace{-7mm}
\end{figure}
As a first application, we evaluate the one-loop cross section of
Bhabha scattering in massive QED. The Bhabha scattering process is
ideally suited to demonstrate the evaluation of processes at NLO by
the Feynman Tree Theorem.  The one-loop result for the cross section
is well known and can easily be produced using automated loop-graph
evaluation tools. Most complications inherent in the method -- UV subtractions, IR cancellations,
and threshold singularities -- are present simultaneously.
There are ten graphs in the one-loop corrections, which after rewriting them
as tree graphs lead to a rich structure in the integrand.  This has to be treated
by multichannel integration methods.  Furthermore, the smallness of the
electron mass compared to the energy of $500\:\GeV$ where we evaluate
the process provides a stringent test of the stability of the numerical integration.

We created analytical expressions for the loop graphs in
computer-readable form using the Mathematica- and FORM-based
packages \feynarts\ and \formcalc~\cite{FeynArts:FormCalc}.
The matrix elements were then handed over to a specially
crafted Mathematica program that creates subtraction
graphs, cuts the loops and, where needed, calculates the fixing
functions. For each tree graph, the program creates parameterizations
(integration channels) which map the resulting phase space onto the
unit hypercube, taking into account the peak structure. The resulting
expressions for the matrix elements and the channels are then written
out in Fortran code.

As an integration routine, we choose the multi-channel algorithm
\vamp~\cite{Ohl:1998jn}.  We compare the final results for the cross
section and angular distribution to an independent calculation that
proceeds along the usual way of integration via tensor reduction, using \feynarts\ and
\formcalc. In the left plot of figure \ref{f_results}, we show the differential
cross section for the full matrix element. The results are in complete
agreement with \feynarts. The adaption of the grids in the multi-channel
approach to the peaks works quite efficient. With 10000 sampling points,
the error estimate on the numerical integration as returned by the Monte-Carlo
integrator is less than $1\%$.

The right hand plot of figure \ref{f_results} shows results for event generation 
of the partonic QED process in the forwards scattering region. The distribution
of events is again in agreement with predictions from \feynarts. Using the
integrand obtained from the FTT, we generate unweighted events with a
phase space including the on-shell momentum of a loop particle. This increases
the dimension of integration from one to four. The efficiency of event
generation is in the percent-regime.

\section{Conclusions}

We reported on a new method for computing NLO corrections to
scattering cross sections. Here, all integrals are transformed into
ordinary phase-space integrals (albeit with unusual boundaries) that
can be handled by an ordinary numerical multi-channel phase-space
integrator.  To this hand, we had not just to implement subtractions
for UV and IR singularities, but furthermore subtraction functions
(fixing functions) for threshold singularities which do not cause
problems in the usual semi-analytic methods.

Extending the method to the full Standard Model and multi-particle
processes, it promises important advantages over more conventional
semi-analytic algorithms: The computational complexity
does not increase dramatically with the number of legs in loop diagrams.
Evaluating a NLO $n$-particle process should require similar CPU
resources as a LO $n+1$-particle process, summed inclusively over all
particle species. Combining loop integration and phase-space sampling
in a single step, we avoid a whole layer in the calculation. In particular,
all terms are evaluated only up to the level of precision that is required by
the actual simulation.

There is still a long way before this method can actually
improve the simulation of physics processes in the Standard Model or
its extensions.  On the one hand, we have to handle the more
complicated IR behavior of QCD and state suitable renormalization
conditions for the non-abelian theory.  On the other hand, the method
has to be augmented by a consistent treatment of unstable states, which
appear in loops and, in our case, would become artificial external particles
in event samples.  Nevertheless, the method, has distinct advantages that
warrant its further development  towards realistic complete LHC and ILC applications.

\end{document}